\documentclass[12pt]{article}
\usepackage{epsfig, amssymb}
\usepackage{graphicx,psfrag} 
\newcommand{\be}{\begin{equation}}
\newcommand{\ee}{\end{equation}}
\newcommand{\bea}{\begin{eqnarray}}
\newcommand{\eea}{\end{eqnarray}}
\newcommand{\bA}{\begin{array}}
\newcommand{\eA}{\end{array}}
\newcommand{\bc}{\begin{center}}
\newcommand{\ec}{\end{center}}
\newcommand{\al}{\alpha}

\newcommand{\ra}{\rightarrow}

\newcommand{\ie}{{\it i.e.}}
\newcommand{\eg}{{\it e.g.}}

\newcommand{\Nf}{${\cal N}{=}4$}
\newcommand{\Nt}{${\cal N}{=}2$}

\newcommand{\vn}{{\vec\nabla}}
\newcommand{\vB}{{\vec B}}
\newcommand{\vE}{{\vec E}}

\newcommand{\vx}{{\vec x}}

\newcommand{\vs}{{\vec s}}

\begin{document}

\begin{titlepage}
\vspace{20mm}

\bc

%\hfill  {TIFR/TH/07-xx} \\
\hfill  {\tt arXiv:0712.3625 [hep-th]} \\
         [22mm]
%%X\vfill

{\Huge On the internal structure of dyons\\ [2mm]
 in \Nf\ super Yang-Mills theories}
\vspace{16mm}

{\large K.~Narayan} \\
\vspace{3mm}
{\small \it Chennai Mathematical Institute, \\}
{\small \it Plot H1, SIPCOT IT Park, \\}
{\small \it Padur PO, Siruseri - 603103, India.\\}
{\small Email: \ narayan@cmi.ac.in}\\

\ec
%\medskip
\vspace{30mm}

\begin{abstract}
We use the low energy effective $U(1)^r$ action on the Coulomb branch 
of ${\cal N}=4$ super Yang-Mills theory to construct approximate field 
configurations for solitonic dyons in these theories, building on the 
brane prong description developed in hep-th/0101114. This dovetails 
closely with the corresponding description of these dyons as string 
webs stretched between D-branes in the transverse space. The resulting 
picture within these approximations shows the internal structure of 
these dyons (for fixed asymptotic charges) to be molecule-like, with 
multiple charge cores held together at equilibrium separations, which 
grow large near lines of marginal stability. Although these techniques 
do not yield a complete solution for the spatial structure (\ie\ all 
core sizes and separations) of large charge multicenter dyons in high 
rank gauge theories, approximate configurations can be found in 
specific regions of moduli space, which become increasingly accurate 
near lines of marginal stability. We also discuss string webs with 
internal faces from this point of view.
\end{abstract}

\end{titlepage}

\newpage 
{\small
\begin{tableofcontents}
\end{tableofcontents}
}

\vspace{2mm}

\section{Introduction}

Understanding the internal structure of solitonic states in string
theories and their low energy limits is an interesting and important
question. In other words, for a localized soliton, given asymptotic
quantum numbers (as seen by a distant observer), we would like to
understand what internal structure one could associate with the
soliton. This is often notoriously difficult in the regimes of most
interest, even in supersymmetric theories. A key system exemplifying
these difficulties is a black hole: given the large Bekenstein-Hawking
entropy, it is tempting to imagine a rich internal structure that is
``fuzzy'' on approximately horizon size. A somewhat simpler but still
fairly rich system in this context constitutes charged solitonic
states in supersymmetric non-abelian gauge theories, which also
helpfully admit realizations in terms of D-brane constructions.

It is well-known that there exist ${1\over 4}$-BPS dyonic states in
\Nf\ $SU(N)$ SYM theories \cite{Bergman:1997yw, Bergman:1998gs,
  Hashimoto:1998zs, Kawano:1998bp, Lee:1998nv, BakLeeYi, Stern:2000ie,
  Gauntlett:1999xz, Kol:2000tw, Ritz:2000xa, Argyres:2001pv,
  Argyres:2000xs, Ritz:2001jk} (see \eg\ \cite{YiRev06} for a recent 
review), represented as string webs \cite{Schwarz96, Dasgupta:1997pu, 
Sen:1997xi} in brane constructions of these
theories. These states can be labelled by their charges w.r.t. the
long range fields, \ie\ the $U(1)^N$ abelian subsector (for
convenience, we regard these as states in $U(N)$ theories Higgsed to
$U(1)^N$ uncharged under an overall $U(1)$). Unlike ${1\over 2}$-BPS
states (whose charges are ``parallel'' or mutually local), the
electromagnetic and scalar forces between the constituent charge
centers of a ${1\over 4}$-BPS state do not precisely cancel except at
specific equilibrium separations (\ie\ there is a nontrivial potential
for the dynamics of the charge cores). There is a rich structure of
the decay of these ${1\over 4}$-BPS states across lines of marginal
stability (LMS) in the moduli space (Coulomb branch) of these
theories. For instance, for a simple 2-center ${1\over 4}$-BPS bound
state, the abelian approximation shows that the spatial size of the 
state (\ie\ the separation between the two charge centers) is 
inversely proportional to the distance from the LMS 
\cite{Ritz:2000xa, Argyres:2001pv, Argyres:2000xs, Ritz:2001jk}. 
Thus as one approaches the LMS, the separation between the centers 
grows and the two charges become more and more loosely bound until 
they finally unbind at the LMS. In terms of the density of states, 
we have a stable 1-particle state on one side of the LMS, which 
decays into the 2-particle state continuum as we cross the LMS.
On the other side of the LMS, the 1-particle state does not exist, and
we only have the 2-particle continuum. The field theory construction
of these \Nf\ string web states dovetails beautifully with
``brane-prong'' generalizations (see \eg\ \cite{Gauntlett:1999xz,
  Argyres:2001pv, Argyres:2000xs}) of the ``brane-spike''
\cite{Callan:1997kz}: these prongs in turn can be interpreted as
string webs stretched between D-branes.

We expect that this picture holds for more general multicenter bound
states too, \ie\ near lines of marginal stability, some of these
states become loosely bound. Then the internal structure in spacetime
of these dyonic states can be described from the point of view of the
low energy abelian $U(1)^N$ effective gauge theory (\ie\ the
nonabelian microscopic physics becomes unimportant) as
``molecule-like'', consisting of several solitonic charge centers
bound together by electromagnetic and scalar forces, qualitatively
somewhat similar to that of the split attractor black hole bound
states of Denef et al \cite{denef, denefHalo} in \Nt\ string theories.
In this paper, we make some modest attempts to make precise these 
general expectations, building on the construction in 
\cite{Argyres:2001pv}.

It is clear that as a function of the moduli values, a typical
solitonic dyon (for fixed asymptotic charges) can have a complicated
internal structure, potentially hard to pin down especially for large
charge in a high rank \Nf\ gauge theory Higgsed to $U(1)^r$ at the
generic point in moduli space. It is therefore interesting to ask if
we can use D-brane constructions to glean insights into the internal
structure of dyons, perhaps focussing on specific regions of moduli
space where abelian approximations are reliable. Towards this end, we
will analyze, in what follows, the spatial molecule-like structure of
these states as it varies on the moduli space, in part using its
limiting description as loosely bound configurations near decay
across lines of marginal stability. While the decay of such a state
into a 2-body endpoint admits an increasingly accurate description
near the LMS, for a general $(n,m)_0$ dyon (where the subscript $0$
refers to some $U(1)$ that these charges refer to), there are several
2-body decay endpoints (corresponding to multiple lines of marginal
stability) depending on how we fix moduli values at charge cores and
how we break up the dyon charges into constituents. Typically each
constituent is itself a composite with further internal structure.
Obtaining a description of such a state turns out to be easier if we
take recourse to the brane construction: starting with a dyon of large
charge which corresponds to a complicated string web, we decompose the
web into smaller sub-webs representing constituent dyons and so on
(sort of reminiscent of wee-partons in a hadron). This typically 
corresponds to a (nested) configuration in spacetime with multiple 
charge cores. The (approximate) internal structure of the parent dyon
resulting from this tree-like process becomes increasingly reliable in
regions of moduli space exhibiting a hierarchy of scales which enables
the constituent dyons (at a given level in the tree) to be pointlike
and widely separated within their parent dyons (immediately above in
the tree). The structure of these states becomes increasingly more
complicated as their charges (or alternatively the number of branes)
increase.

We then use these techniques to study the internal structure of dyons 
corresponding to string webs with internal faces.

We first review $SU(3)$ dyons/webs in Sec.~2, and then describe
transitions in the dyon internal structure as we move on the moduli
space in Sec.~3. In Sec.~4.1, we describe $SU(4)$ dyon/web
configurations, while Sec.~4.2, 4.3 describe more general dyon/web
states of higher charge. Sec.~5 discusses webs with internal faces.
Finally we discuss a few issues in Sec.~6, in particular pertaining to
how reliable these configurations are.  An Appendix reviews the basic
framework we use here.

\section{String webs from field theory}\label{sec:0101114}

We give here a brief description of the construction of string web
dyon states in $SU(N)$ SYM theories from their low energy $U(1)^N$
(Higgsed from $U(N)$, with an overall decoupled $U(1)$) effective
theories, in part making transparent their corresponding D3-brane
constructions (this mainly follows \cite{Argyres:2001pv,
  Argyres:2000xs}). The general field theory strategy is to extremize
the energy functional and construct $2N$ first-order Bogomolny
equations which relate the electric and magnetic fields linearly to
gradients of the scalar fields, which can then be solved subject to
certain charged source boundary conditions, such that the resulting
solutions extremize the mass of the charged states in question. The 
scalar field configurations obtained thus are maps\ $(X_i(\vs),Y_i(\vs)), 
i=1,\ldots,N$, from spacetime to the moduli space of the gauge theory. 
These approximate solutions to the $U(1)^N$ theory become more and more
exact in the vicinity of lines of marginal stability and give a 
constructive answer to the question of the existence and stability 
of these states. They describe string webs on the moduli space of 
the field theory, which can then be shown to ``fold'' into
string webs stretching between D-branes in transverse space. The
minimax problem involved in the field theory construction is
straightforward but not simple, especially for higher rank
theories. However in known examples, it effectively reproduces
brane-prong configurations (generalizations of ${1\over 2}$-BPS
brane-spikes \cite{Callan:1997kz}) that can be written down relatively
simply and intuitively: it is these effective brane-prong field theory
configurations that we will find useful here.

\subsection{$SU(3)$ dyons and webs: mostly a review}

For simplicity and ease of illustration, we describe ${1\over
  4}$-BPS string web dyon states in $SU(3)$ SYM theory (along the
lines discussed at length in \cite{Argyres:2001pv, Argyres:2000xs}, 
and reviewed briefly in Appendix A), arising as the low energy theory 
on three non-coincident D3-branes, Higgsing $U(3)\ra U(1)^3$ (with 
an overall decoupled $U(1)$). This is mainly a review (but presented 
slightly differently from \cite{Argyres:2001pv}), meant to set our 
notation for what follows.

Let us first recall that ${1\over 2}$-BPS states are only charged with 
respect to a single (relative) $U(1)$. Thus the electric/magnetic 
charge vectors of a ${1\over 2}$-BPS charge $(p,q)$ state are
\be
Q_e=p\alpha=p(e_1-e_2)\ , \qquad Q_m=q\alpha=q(e_1-e_2)\ ,
\ee
where we have defined simple roots $\alpha=e_1-e_2, \beta=e_2-e_3$, with 
$\alpha^2=\beta^2=2, \alpha\cdot\beta=-1$, and $e_i, i=1,2,3$ are the 
(orthonormal) roots of $U(1)^3$, with $e_i\cdot e_j = \delta_{ij}$.
There of course exist ${1\over 2}$-BPS states charged under different 
$U(1)$s as well. Labelling these states by their charges w.r.t. $U(1)^2$ 
(the total charge is zero, decoupling the overall $U(1)$), and writing
the charge vectors in terms of the $e_i$ basis of $U(1)^3$ makes
transparent the connection to the brane constructions of these states. 
For example, the above ${1\over 2}$-BPS state can be interpreted as a
$(p,q)$ (oriented) string stretched between two D3-branes, with the
two D3-branes carrying point-like dyons of charge $(p,q)$ and
$(-p,-q)$ respectively. As an example, the scalar field configuration 
(for the two scalars of the two D-branes) representing say an electric 
charge $(1,0)$ in a $U(1)^2$ theory is
\be\label{spikesU12}
X_1 = {e\over |\vs-\vs_0|} - X_0\ , \qquad X_2 = -{e\over |\vs-\vs_0|} + L\ ,
\ee
where $e$ is the unit of electric charge, and the two D3-branes are 
located at $X=-X_0$ and $X=L$. For magnetic charges, we have say 
$X_1={g\over|\vs-\vs_0|}-X_0$, with $g$ the unit of magnetic charge. 
The two separate brane spikes join at $\vs\ra\vs_0$, where $X_1, 
X_2\ra X_0'$:\ regulating the divergence in this abelian 
approximation\footnote{Note that these are approximate solutions in this 
abelian framework: \eg\ the two sides $X_1(\vs)$ and $X_2(\vs)$ do not 
join smoothly. Note also that the Bogomolny bound equations for the 
BPS sector from the full nonlinear Born-Infeld action are the same as 
the ones from this leading order approximation (although their masses 
might differ by numerical factors). That these effective actions are 
insufficient is not surprising, since near a charge core, the field 
strengths are not slowly varying. In this region, new physics (higher 
derivative contributions, nonabelian physics etc) enters.}, this gives
\be
e\al_{\vs_0,\vs_0}-X_0=X_0'=-e\al_{\vs_0,\vs_0}+L \qquad\ 
\Rightarrow \qquad\  e\al_{\vs_0,\vs_0} = {L+X_0\over 2}\ ,
\ee
where $\al_{\vs_0,\vs_0}$ is the (approximate) inverse core size. 
This gives the location of gluing on the moduli space as\ 
$X_0'={L-X_0\over 2}$, which is the midpoint of the line joining the 
two branes, corresponding to the singularity of enhanced $U(2)$ gauge 
symmetry. For the magnetic charge or monopole, this agrees with our 
expectation that the nonabelian fields are nontrivial inside the 
monopole core (roughly given by the Higgs vev), which via the scalar 
configurations is located at $(X,Y)=(X_0',0)$ on the moduli space. In 
the abelian description here, these are approximated as Dirac monopoles.

Analysing the energy\footnote{A field theory vev $X$ and its 
corresponding coordinate length in transverse space $x$ are related as 
$X={x\over\al'}$, as can be seen from \eg\ the D-brane DBI action.} of 
this state to compare with the mass\ ${1\over\al'}(l+x_0)=(L+X_0)$ of 
a fundamental string of tension ${1\over\al'}$ stretched between the 
D-branes (and a similar analysis for a D-string) shows that we must set 
\be
e=g_s, \qquad g=1\ .
\ee
(the numerical factors are not important for what follows.)
This makes intuitive sense: at weak coupling (small $g_s=g_{YM}^2$), we 
see as expected that the fundamental string ending on the brane at 
$\vs=\vs_0$ causes only a small deformation to the brane worldvolume 
(as shown by say $X_1$) away from $\vs_0$, where we see a sharp spike. 
However the D-string is not a small perturbation as reflected by the
$g_s$-independence of the magnetic charge unit $g$. In spacetime, this 
implies that the monopole core size is $O({1\over L+X_0})$, \ie\ set by 
the Higgs expectation value, while the electric charge core size is 
$O({g_s\over L+X_0})$: thus for small $g_s$, electric charge cores can 
be regarded as pointlike.

\begin{figure}
\bc
\epsfig{file=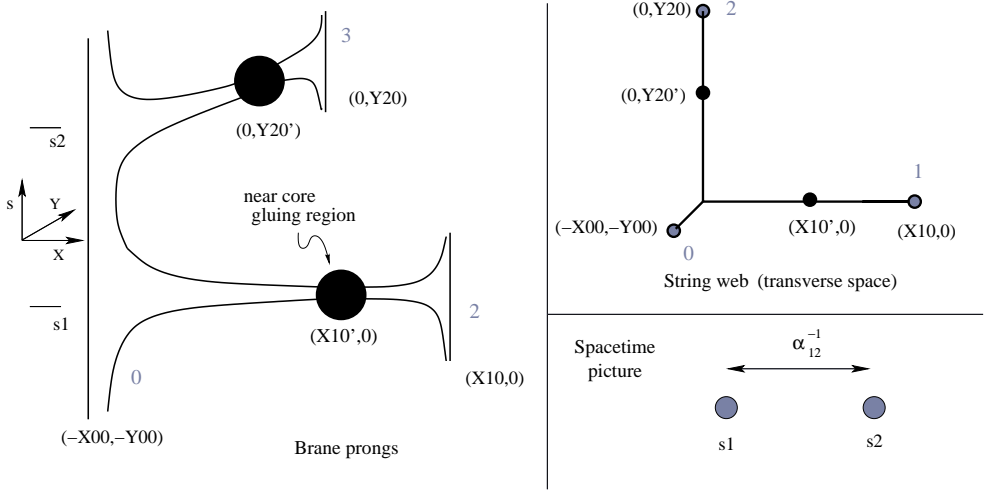, width=13cm}
\caption{{\small Field theory brane-prong construction of an $SU(3)$ 
string web. $\vs$ parametrizes worldvolume/spacetime. The grey circles 
are the D-brane locations while the black circles show the locations 
where the brane-prongs glue onto each other at the charge cores 
$\vs_1,\vs_2$.}}
\label{webprongs}
\ec
\end{figure}

In contrast to ${1\over 2}$-BPS states, ${1\over 4}$-BPS states in 
$SU(3)$ SYM theory are charged under both $U(1)$s. The electric and 
magnetic charge vectors of the generic $SU(3)$ web labelled as 
$[(p_1+p_2, q_1+q_2), (-p_1,-q_1), (-p_2,-q_2)]$ are
\bea\label{QeQmSU3}
&& Q_e = p_1.\alpha + p_2.(\alpha+\beta) 
= (p_1+p_2)e_1 + (-p_1)e_2 + (-p_2)e_3\ , \nonumber\\
&& Q_m = q_1.\alpha + q_2.(\alpha+\beta) 
= (q_1+q_2)e_1 + (-q_1)e_2 + (-q_2)e_3 \ .
\eea
From the expressions in terms of the $e_i$s, it is straightforward to 
interpret this as a string web (see Figure~\ref{webprongs}), with  
$(p_1+p_2, q_1+q_2), (p_1,q_1)$ and $(p_2,q_2)$ strings constituting 
the three legs ending on three D3-branes. Regarding outgoing charge 
from a charge center as positive, the dyon charges on the three 
D3-brane worldvolumes are $(p_1+p_2, q_1+q_2), (-p_1,-q_1)$ and 
$(-p_2,-q_2)$. 

The field theory description shows that one can think of this state as
a charge $(p_1+p_2, q_1+q_2)$ dyon, spatially consisting of two
constituent charge centers $(p_1,q_1), (p_2,q_2)$, separated by a
distance inversely proportional to the distance from the LMS. Then as 
one approaches the LMS, the ${1\over 4}$-BPS state (represented as 
$(Q_e,Q_m)$) decays as
\be
(p_1.\alpha + p_2.(\alpha+\beta),\ q_1.\alpha + q_2.(\alpha+\beta))\ \ 
\ra\ \ (p_1\alpha,\ q_1\alpha)\ +\ (p_2(\alpha+\beta),\ q_2(\alpha+\beta))\ ,
\ee
into the two constituent ${1\over 2}$-BPS dyons of charge $(p_1,q_1)$ and 
$(p_2,q_2)$ representing the two separate ${1\over 2}$-BPS $(p_1,q_1)$ and 
$(p_2,q_2)$ strings. Here we describe the simplest such configuration\ 
$(1,1)-(-1,0)-(0,-1)$. The configuration of scalars $X_i,Y_i,\ i=0,1,2$ 
describing this string web (Figure~\ref{webprongs}) is
\bea\label{111001web}
X_0 = {e\over |\vs-\vs_1|} - X_0^0\ , && \qquad
Y_0 = {g\over |\vs-\vs_2|} - Y_0^0\ , \nonumber\\
X_1 = -{e\over |\vs-\vs_1|} + X_1^0\ , \qquad Y_1=0\ , \quad && \quad 
X_2=0\ , \qquad Y_2 = -{g\over |\vs-\vs_2|} + Y_2^0\ .
\eea
From the D-brane point of view, the interpretation is clear (see
Figure~\ref{webprongs}): $X_i,Y_i$ represent scalars of the three
D-branes $i=0,1,2$, with worldvolumes parametrized by $\vs$, so that
this configuration $(X_i(\vs),Y_i(\vs))$ describes brane-prong
deformations of the three brane worldvolumes.  We have the boundary
conditions on the scalars:
\bea
&& \vs\ra\infty:\ (X_0,Y_0)\ra (-X_0^0,-Y_0^0)\ ,\quad 
(X_1,Y_1)\ra (X_1^0,0)\ , \quad (X_2,Y_2)\ra (0,Y_2^0)\ , \nonumber\\
&& \vs\ra\vs_1:\ (X_0,Y_0)\ra ({X_1^0}',0) \leftarrow (X_1,Y_1)\ , \nonumber\\
&& \vs\ra\vs_2:\ (X_0,Y_0)\ra (0,{Y_2^0}') \leftarrow (X_2,Y_2)\ ,
\eea
Defining the inverse core separations and core sizes $\al_{ij}\geq 0$
\be
\al_{ij}\equiv {1\over |\vs_i-\vs_j|}\ , \ \ i\neq j, \qquad 
\al_{ii}={1\over\epsilon_i}\ ,
\ee
for some cutoffs $\epsilon_i$ on the charge core sizes, we obtain the 
following constraints on the existence of a solution to the above system 
within the $U(1)^N$ approximation:
\bea\label{111001bndrycondns}
&& \vs\ra\vs_1:\quad {X_1^0}' = e\al_{11} - X_0^0 = -e\al_{11} + L_1\ , 
\quad\ 0 = g\al_{12} - Y_0^0 \ , \nonumber\\
&& \vs\ra\vs_2:\quad 0 = e\al_{12} - X_0^0\ , \quad\ 
{Y_2^0}' = g\al_{22} - Y_0^0 = -g\al_{22} + Y_2^0\ ,
\eea
giving
\bea\label{111001cores}
{X_1^0}' = {X_1^0 - X_0^0\over 2}\ , 
& \qquad & {Y_2^0}' = {Y_2^0 - Y_0^0\over 2}\ , \nonumber\\
e\al_{11} = X_1^0-{X_1^0}' = {X_1^0+X_0^0\over 2}\ , \ \
& g\al_{22}=& Y_2^0-{Y_2^0}' = {Y_2^0+Y_0^0\over 2}\ ,
\ \ \ e\al_{12} = X_0^0\ .
\eea
This thus solves for the charge core sizes $\al_{ii}^{-1}$ and the 
separation between the charge core centers 
\be
r_{12}={1\over\al_{12}}\sim {g_s\over X_0^0}\ ,
\ee 
and further gives also the constraint (since $\al_{12}$ appears in both 
lines of (\ref{111001bndrycondns}))
\be
{e\over g} Y_0^0 = X_0^0\ .
\ee
Clearly a solution with these charges exists only if the physical 
core separation $r_{12}>0$, \ie\ only for $X_0^0,Y_0^0>0$, \ie\ on one 
side of the line of marginal stability, which passes through the 
junction at $(X,Y)=(0,0)$. Note that $r_{12}$ is inversely proportional 
to the length of the shortest leg (\ie\ the $(1,1)$-string) of the 
string web.
As $X_0^0\ra 0$, we see that $\al_{12}\gg \al_{ii}$, \ie\ we have two 
widely separated pointlike charge cores. In more detail, the abelian 
approximation is good when\ $r_{12}\gg r_{11},r_{22}$, \ie\ in the region 
of moduli space where\ $X_0^0\ll X_1^0$\ and\ $X_0^0\ll {e\over g}Y_2^0$.
In this region, the $U(1)^3$ approximation 
neglecting the microscopic nonabelian physics of the charge cores
becomes increasingly good, and this molecule-like description in terms
of pointlike dyon constituents is reliable. Furthermore, the single
$U(1)$ from brane-0 with only charge boundary conditions from
branes-1,2 yields an increasingly good description of the spatial
structure of the dyon-web (see Figure~\ref{webprongs}).

In terms of the brane-prong interpretation, we see that the $\vs_1$
leg of the prong from brane-1 joins the corresponding leg from brane-2
at the location $({X_1^0-X_0^0\over 2},0)$ of the charge core $\vs_1$,
while the $\vs_2$ leg of the prong from brane-1 joins the
corresponding leg from brane-3 at the location $(0,{Y_2^0-Y_0^0\over 2})$ 
of the charge core $\vs_2$. Note that the existence and structure of 
the dyon configuration is closely tied to the geometry of the D-branes 
in transverse space, \eg\ $\al_{11}>0 \Longrightarrow\ X_1^0>{X_1^0}'$ 
and so on.

The cutoff sizes (${1\over\al_{ii}}$) also shows that the charge cores 
are located at the enhanced $SU(2)$ symmetry points 
$({X_1^0}',0)=({X_1^0-X_0^0\over 2},0)={1\over 2}((X_1^0,0)+(-X_0^0,0))$ 
and $(0,{Y_2^0}')=(0,{Y_2^0-Y_0^0\over 2})$ on the $SU(3)$ moduli space. 
At these locations, there are light nonabelian fields which must be 
included into the low energy effective action for a nonsingular 
description. Clearly, there is also a $symmetric$ $point$ 
$X_1^0=X_0^0={e\over g}Y_0^0={e\over g}Y_2^0$ in moduli space where 
the core sizes are comparable to their relative separations so that 
the charge centers effectively coalesce.

\section{$SU(3)$ dyons: moduli space transitions in internal structure}

Now we discuss what seems to be a generic feature of these field
theory dyons: this is the fact that the apparent internal structure of
these $SU(3)$ dyons undergoes transitions as we move around on the
moduli space.  It is easiest to describe this in a specific example:
consider the $(1,1)-(-1,0)-(0,-1)$ string web. Keeping explicitly the 
charges w.r.t. all of the $U(1)$s in the theory, the asymptotic charges 
of this dyon at spatial infinity can be read off from the charge 
vectors $(Q_e,Q_m)$ in (\ref{QeQmSU3}) as
\be
\{\ (1,1)_0\ (-1,0)_1\ (0,-1)_2\ \}\ ,
\ee
where the subscript labels the $U(1)$ that the charge values refer to.
\begin{figure}
\bc
\epsfig{file=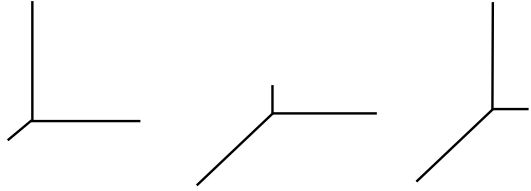, width=7cm}
\caption{{\small The shape of the $SU(3)$ string web in the three 
regions of moduli space, with the moduli values as shown in 
Figure~\ref{webprongs}.}}
\label{SU3modsptrans}
\ec
\end{figure}

Consider the region in moduli space where $X_0^0,Y_0^0\ll X_1^0,Y_2^0$:\ 
here the $(1,1)$ leg of the web is short and the corresponding field
configuration is described in (\ref{111001web}), with the core sizes 
and separations given in (\ref{111001cores}). As we have seen, the
internal structure of the dyon in this region can be essentially
thought of as consisting of two charge cores of charges
\be
X_0^0,Y_0^0\ll X_1^0, Y_2^0: \qquad\ 
\{(1,0)_0 (-1,0)_1\}\ , \quad \{(0,1)_0 (0,-1)_2\}\ .
\ee
Now consider the region in moduli space where $Y_2^0\ll X_0^0,Y_0^0,X_1^0$, 
\ie\ the $(0,1)$ leg of the web is short. It is straightforward to study 
the spatial structure from constructing a similar field configuration as 
before. We then find that the dyon can now be thought of as consisting 
of two cores of charges
\be
Y_2^0\ll X_0^0,Y_0^0,X_1^0: \qquad\ 
\{(-1,-1)_2 (1,1)_0\}\ , \quad \{(1,0)_2 (-1,0)_1\}\ .
\ee
Similarly, the region\ $X_1^0\ll X_0^0,Y_0^0,Y_2^0$\ in moduli space 
where the $(1,0)$ leg of the web is short shows the dyon constituents 
to have charges
\be
X_1^0\ll X_0^0,Y_0^0,Y_2^0: \qquad\
\{(-1,-1)_1 (1,1)_0\}\ , \quad \{(0,1)_1 (0,-1)_2\}\ .
\ee

In going from one of these regions to another in moduli space, one does 
not cross any line of marginal stability: clearly the dyon exists and 
is stable in these transitions. Furthermore it is clear that there is 
no violation of charge conservation since the charges at spatial 
infinity are unchanged throughout. What is happening in the process 
of transiting between any two of these regions in moduli space is simply 
charge rearrangement. In the region where say the web leg ending on 
D-brane $a$ is shortest, \ie\ where D-brane $a$ is closest to the LMS, 
the charges w.r.t. this $U(1)_a$ break up into constituents, and 
similarly in the other regions of moduli space\footnote{It is worth 
recalling something similar in \eg\ Seiberg-Witten theory \cite{sw94}: 
the W-boson of charge $(2,0)$, elementary at weak coupling, is best 
described as a string-web-like configuration (in the D-brane 
construction via F-theory \cite{swFth}), a bound state of the 
monopole (charge $(0,1)$) and dyon (charge $(2,-1)$), that decays 
across the line of marginal stability in the strong coupling region. 
A description of this decay in spacetime appears in \eg\ 
\cite{Argyres:2001pv} (using equilibrium scalar configurations as 
here) and \cite{Ritz:2001jk} (using the long-range forces between 
the constituents).}.

Let us look more closely at the transition between say the region
$X_0^0,Y_0^0\ll X_1^0, Y_2^0$, and $X_1^0\ll X_0^0,Y_0^0,Y_2^0$. From
the web, it is clear that we extend the $(1,1)$ leg and shrink the
$(1,0)$ leg, while correspondingly in spacetime, the size
${1\over\al_{11}}$ of one of the cores increases with the core
separation ${1\over\al_{12}}$ decreasing. At some intermediate point
$X_1^0=X_0^0$, we have\ ${1\over\al_{11}}={1\over\al_{12}}$:\ from 
(\ref{111001cores}), we see that this point, where ${X_1^0}'=0$,
corresponds to the singularity of enhanced $SU(2)$ symmetry on the
$SU(3)$ moduli space. This is not surprising since we expect that the
nonabelian degrees of freedom become light and cannot be neglected 
when the dyon constituents approach each other. The Higgs vev 
hierarchies are different on either side of the transition.

Analyzing the other transitions shows similar structure: transitions in 
the internal structure pass through singularities of enhanced symmetry 
in the moduli space.

\section{$SU(N)$: general $(n,m)$ dyon web states}

Now let us try to understand the internal structure of more general 
dyons: for concreteness, we consider a dyon whose charge w.r.t. some 
$U(1)$ is $(n,m)_0$  (where the subscript $0$ refers to the $U(1)_0$ 
w.r.t. which this charge is defined), and study the corresponding field 
configuration, which should then give insights into the internal 
structure.

This state (unless ${1\over 2}$-BPS), will typically exhibit a rich 
structure of decay across one or more lines of marginal stability, where 
it becomes loosely bound.
The field configuration describing this can be written down reliably in 
the vicinity of the simultaneous location (coincidence) of the various 
lines of marginal stability at which this state can decay. A simple and 
intuitive way to obtain the field configuration without ``deriving'' it 
rigorously is to note that ultimately this state comprises and therefore 
decays into $n$ $(1,0)$ strings and $m$ $(0,1)$ strings. Far from any 
LMS, we expect that the $(n,m)_0$ dyon is pointlike so that the
part of the scalar field configuration stemming from $U(1)_0$ is
\be
X_0 = {n e\over |\vs-\vs_0|} - X_0^0\ , \qquad \
Y_0 = {m g\over |\vs-\vs_0|} - Y_0^0\ , 
\ee
\ie\ a single spike emanating from $\vs=\vs_0$, representing the 
$(n,m)$ string beginning on a D-brane located at $(-X_0^0,-Y_0^0)$, and 
ending on a stack of D-branes located at approximately $(X_{av}^0,Y_{av}^0)$. 
However as we move on the moduli space to split up the D-brane stack
at $(X_{av}^0,Y_{av}^0)$ (on the Coulomb branch), the $(n,m)_0$ dyon
constituents break up and begin to separate, revealing some internal
structure. In other words, the charge core at $\vs_0$ gets resolved into 
multiple distinct charge cores within as the single D-brane stack at
$(X_{av}^0,Y_{av}^0)$ splits to a multicenter solution on the Coulomb
branch. Each core is charged w.r.t. $U(1)_0$ as well as one or more 
other $U(1)$s: this corresponds to the fact that each constituent 
string (or string web) is stretched between two or more D-branes, 
thus carrying charge under the corresponding $U(1)$s. Clearly we can 
split up the D-brane stack in many distinct ways: this leads to 
correspondingly distinct internal structures for the $(n,m)_0$ dyon, 
depending on how we split up the $(n,m)$ charge.

As an example, consider say a dyon of charge $(5,3)_0$. With no
splitting, \ie\ with a single stack at $(X_{av}^0,Y_{av}^0)$, we have a
${1\over 2}$-BPS state in effectively an $SU(2)$ theory, corresponding
to the $(5,3)$ string stretched between the D-branes. Now split the 
$(X_{av}^0,Y_{av}^0)$ stack into two: depending on how we break up the 
$(5,3)$ charge into two, we get different string webs with three legs 
in an $SU(3)$ theory.
Similarly starting with the $(n,m)$ dyon/string emanating from the 
D-brane at $(-X_0^0,-Y_0^0)$, and splitting up the $(X_{av}^0,Y_{av}^0)$ 
stack into say $k$ centers on the Coulomb branch gives distinct string 
webs with $k+1$ legs in an $SU(k+1)$ theory, depending on how the 
$(n,m)$ charge is broken up into constituents. In other words, for 
each leg of the string web in transverse space, we have a charge core 
in spacetime. Assuming the final elementary constituents to be only 
$(1,0)$ and $(0,1)$ charges, we can split a $(n,m)$ dyon/string into 
irreducible string webs at most in an \Nf\ $SU(n+m+1)$ gauge theory.

Such a maximally split $(n,m)_0$ dyon state corresponds to the 
charge vectors
\be
Q_e = ne_0 + \sum_{i=1}^n (-1)e_{E_i}\ , \qquad
Q_m = me_0 + \sum_{i=1}^m (-1)e_{M_i}\ ,
\ee
in an \Nf\ $SU(n+m+1)$ gauge theory\footnote{These can also be regarded 
as dyonic states studied by Stern and Yi \cite{Stern:2000ie} (in the 
context of identifying their degeneracies), given by charge vectors
\bea
Q_m = \sum_i \alpha_i\ , \quad 
Q_e = (\nu+\sum_jq_j)\alpha_1 + (\nu-q_1+q_2+\ldots)\alpha_2 + 
(\nu-q_1-q_2+\ldots)\alpha_3 + \ldots + (\nu-\sum_jq_j)\alpha_{N-1}\ , 
\nonumber
\eea
(for appropriate $\nu,q_i$) or equivalently, using the roots 
$\al_1=e_1-e_2 ,\ \al_2=e_2-e_3 ,\ \ldots$,\ of $SU(N)$, 
\bea
Q_m = e_1-e_N\ , \qquad 
Q_e = (\nu+\sum_iq_i)e_1 + \sum_{i=2}^{N-2}(-2q_i)e_i 
+ (-\nu+\sum_iq_i)e_N\ . \nonumber
\eea
}.
The endpoint charges at each $e_k$ show that this state represents 
a string web with $n+m+1$ legs carrying charges $$\{ (n,m)_0, (-1,0)_{E_1}, 
(-1,0)_{E_2}, \ldots, (-1,0)_{E_n}, (0,-1)_{M_1}, \ldots, (0,-1)_{M_m} \},$$ 
and ending on the $n+m+1$ D-branes ($e_0$ represents $U(1)_0$). 
Diagrammatically this can be represented as the web in
Figure~\ref{nmweb} (the Figure shows the $(5,3)$ web described in more
detail later). This state is classically BPS since its constituents
are essentially $(1,0)$ and $(0,1)$ strings which together preserve a
${1\over 4}$-th of the supersymmetry (see however the Discussion,
Sec.~6, for more on this). This is vindicated by the field
configuration we exhibit below as a solution to first order Bogomolny
bound equations in the SYM theory. For $m=1$, this reduces to a
$(n,1)_0$ dyon, \ie\ a monopole with $n$ electric charges attached:
this is clearly the simplest such dyonic state and there are
simplifications in its internal structure. It is thus efficient to
glean insights into dyons of higher magnetic charge by splitting their
charges to ultimately reduce to the form of an $(n,1)_0$ state. A
systematic way to implement this is obtained by noticing the
sequential decomposition
\be
(n,m)\ra\ (n,m-1) + (0,1) \ra\ (n,m-2) + (0,1) + (0,1) \ra\ \ldots\ (n,1) + 
\sum^m (0,1)
\ee
of the $(n,m)_0$ state. This is of course reliable in specific regions 
of moduli space, which we will describe below. Also note that this 
decomposition gives non-degenerate constituents only if $n,m$ are prime 
(not just coprime).

The D-brane worldvolume scalars describing the string web representing 
this $(n,m)_0$ dyonic state in this $SU(n+m+1)$ theory can then be 
written, generalizing the simple 2-center $SU(3)$ web (\ref{111001web}), 
as 
\bea\label{nmEqns}
X_0 = \sum_{i=1}^{E_n} {e\over |\vs-\vs_i|} - X_0^0\ , && \qquad
Y_0 = \sum_{i=1}^{M_m} {g\over |\vs-\vs_i|} - Y_0^0\ , \nonumber\\
X_k = -{e\over |\vs-\vs_k|} + X_k^0\ , && \qquad
Y_k = Y_k^0\ , \qquad\qquad\qquad\ \ k=E_1,\ldots,E_n, \nonumber\\
X_k = X_k^0\ , && \qquad  Y_k = -{g\over |\vs-\vs_k|} + Y_k^0\ , 
\quad k=M_1,\ldots,M_m\ . \ \
\eea
This is valid in the region of moduli space where brane-$0$ is near one 
or more lines of marginal stability. The intuition for writing these 
field configurations is as we have mentioned above: for brane-$0$, the 
scalars $X_0,Y_0$ have prongs extending from each of the charge cores, 
which then glue onto single charge spikes from other branes-$k$.
The boundary conditions on these scalars are
\bea\label{nmBndrycondns}
\vs\ra\infty: && (X_0,Y_0)\ra (-X_0^0,-Y_0^0)\ ,\qquad 
(X_k,Y_k)\ra (X_k^0,Y_k^0)\ , \quad k\in \{E_i, M_j\}\ ,
\nonumber\\
\vs\ra\vs_{E_k}: && (X_0,Y_0)\ra\ ({X_{E_k}^0}',{Y_{E_k}^0}') \leftarrow\
(X_{E_k},Y_{E_k})\ , \nonumber\\
\vs\ra\vs_{M_k}: && (X_0,Y_0)\ra\ ({X_{M_k}^0}',{Y_{M_k}^0}') \leftarrow\
(X_{M_k},Y_{M_k})\ ,
\eea
the $({X_k^0}',{Y_k^0}')$ being the locations in transverse space 
where the prongs glue onto each other\ (to be distinguished from 
the vacuum moduli values $(X_k^0,Y_k^0)$).
These then give constraint equations on the field configuration to exist:
\bea\label{nmConstraints}
\vs\ra\vs_{E_k}: && {X_{E_k}^0}' = e\sum_{i=1}^{E_n}\al_{E_k,E_i}-X_0^0 
= -e\al_{E_k,E_k}+X_k^0\ , \nonumber\\
&& {Y_{E_k}^0}' = g\sum_{i=1}^{M_m}\al_{E_k,M_i}-Y_0^0 = Y_k^0\ , \nonumber\\
\vs\ra\vs_{M_k}: && {X_{M_k}^0}' = e\sum_{i=1}^{E_n}\al_{M_k,E_i}-X_0^0 
= X_k^0\ , \nonumber\\
&& {Y_{M_k}^0}' = g\sum_{i=1}^{M_m}\al_{M_k,M_i}-Y_0^0 
= -g\al_{M_k,M_k}+Y_k^0\ .
\eea

In what follows, we use the above equations to write out scalar field
configurations and the corresponding constraints from boundary
conditions for dyons in various \Nf\ theories, and glean insights into
their internal structure.

\subsection{$SU(4)$ dyons: 3-center configurations}

\begin{figure}
\bc
\epsfig{file=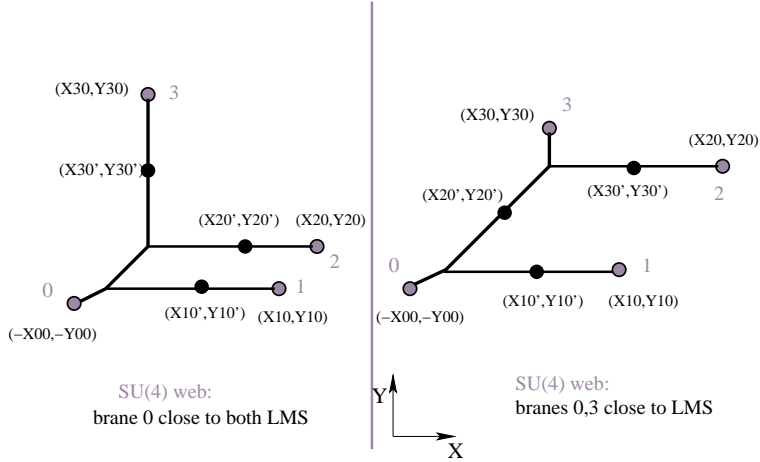, width=10cm}
\caption{{\small String webs in an $SU(4)$ theory: shapes in two 
different regions of moduli space. Grey circles: D-branes, black 
circles: prong gluing locations.}}
\label{websSU4}
\ec
\end{figure}

In this section, we study string web states describing 3-center dyon
bound states in the $U(1)^3$ Higgsed theory stemming from $SU(4)$ (or
more precisely $U(4)\ra U(1)^4$). Consider the $(2,1)_0$ dyon, 
represented by the string web $(2,1)-(-1,0)-(-1,0)-(0,-1)$. This can 
be described by the following 3-center scalar configuration (left 
side of Figure~\ref{websSU4}):
\bea
X_0 = {e\over |\vs-\vs_1|} + {e\over |\vs-\vs_2|} - X_0^0\ , && \quad 
Y_0 = {g\over |\vs-\vs_3|} - Y_0^0\ , \nonumber\\
X_1 = -{e\over |\vs-\vs_1|} + X_1^0\ , \quad Y_1=Y_1^0\ , \qquad\ &&
X_2 = -{e\over |\vs-\vs_2|} + X_2^0\ , \quad Y_2=Y_2^0\ , \nonumber\\
X_3 = X_3^0\ , \quad && Y_3 = -{g\over |\vs-\vs_3|} + Y_3^0\ .
\eea
In the limit where all centers coalesce, \ie\ $\vs_1,\vs_2,\vs_3\sim\vs_0$, 
the deformation from brane-0 resembles a spike representing a 
$(2,1)$-string. Thus the $\vs_1,\vs_2$ centers are electric charges, 
while $\vs_3$ is a monopole. We have the boundary conditions near each 
charge core:
\bea
\vs\ra\infty: && (X_0,Y_0)\ra (-X_0^0,-Y_0^0)\ ,\qquad 
(X_k,Y_k)\ra (X_k^0,Y_k^0)\ , \nonumber\\
\vs\ra\vs_{1,2}: && (X_0,Y_0)\ra\ ({X_{1,2}^0}',{Y_{1,2}^0}') \leftarrow\
(X_{1,2},Y_{1,2})\ , \nonumber\\
\vs\ra\vs_3: && (X_0,Y_0)\ra\ ({X_3^0}',{Y_3^0}') \leftarrow\ (X_3,Y_3)\ .
\eea
This gives the constraints:
\bea
\vs\ra\vs_1:&& \quad {X_1^0}'=e\al_{11}+e\al_{12}-X_0^0=-e\al_{11}+X_1^0\ ,
\qquad\ {Y_1^0}'=g\al_{13}-Y_0^0=Y_1^0\ , \nonumber\\
\vs\ra\vs_2: && \quad {X_2^0}'=e\al_{12}+e\al_{22}-X_0^0=-e\al_{22}+X_2^0\ , 
\qquad\ {Y_2^0}'=g\al_{23}-Y_0^0=Y_2^0\ , \nonumber\\
\vs\ra\vs_3:&& \quad {X_3^0}'=e\al_{13}+e\al_{23}-X_0^0=X_3^0\ , 
\quad\ {Y_3^0}'=g\al_{33}-Y_0^0=-g\al_{33}+Y_3^0\ .
\eea
It is straightforward to simplify these equations towards solving for 
the core sizes and separations, and we find in particular that the 
electric-magnetic core separations are fixed as
\be
g\al_{13}=Y_0^0+Y_1^0\ , \qquad g\al_{23}=Y_0^0+Y_2^0\ ,
\ee
showing the two distinct lines of marginal stability to be at 
$\al_{13}\ra 0$ and $\al_{23}\ra 0$, \ie\ at $Y_0^0,Y_1^0\ra 0$, and 
$Y_0^0,Y_2^0\ra 0$. 
Note however that the electric-electric core separations are not
determined completely. In this case, with a single monopole, the
magnetic core size is also fixed, while the electric core sizes are
fixed only upto the separations. This incompleteness in the solution
turns out to be a generic feature as we go to higher charge, as we
will see in what follows. Finally, consistency of the solution
(compatibility of the ${Y_k^0}', {X_3^0}'$ equations above) gives the
constraint
\be\label{modconst21SU4}
{e\over g} (2Y_0^0+Y_1^0+Y_2^0)=X_0^0+X_3^0\ ,
\ee
on the moduli values of this solution. To elucidate the meaning of 
this constraint, notice that for large $X_0^0,Y_0^0$, this simplifies 
to ${e\over g} 2Y_0^0=X_0^0$, which describes the line of slope 
${y\over x}\sim{1\over 2}$ traced out by the $(2,1)$-string in the 
transverse space, while the limit $Y_0^0,X_0^0$ and $Y_1^0\ra 0$ 
gives ${e\over g}Y_2^0=X_3^0$, which describes the intermediate 
$(1,1)$-string leg between the two junctions in the string web.

Note that it is in the vicinity of the simultaneous coincidence of
both LMS (\ie\ short $(2,1)$- and intermediate $(1,1)$- legs; left
side of Figure~\ref{websSU4}) that the above field configuration can
be regarded as reliable: in this region of moduli space
$X_0^0,Y_0^0,Y_1^0,Y_2^0\ra 0$, the core sizes are\ 
${1\over\al_{ii}}\sim {1\over X_1^0},\ {1\over X_2^0},\ {1\over Y_3^0}$, 
all effectively pointlike for large vevs, with the electric charges 
widely separated from the monopole.

The asymptotic charges at spatial infinity and those of the individual 
centers are
\bea
\{\ (2,1)_0\ (-1,0)_1\ (-1,0)_2\ (0,-1)_3\ \} \qquad\ 
{\rm at\ spatial\ infinity}\ ,\nonumber\\
\vs_1\equiv \{(1,0)_0 (-1,0)_1\}\ , \quad 
\vs_2\equiv \{(1,0)_0 (-1,0)_2\}\ , \quad 
\vs_3\equiv \{(0,1)_0 (0,-1)_3\}\ .
\eea
It is straightforward to check that there are transitions in the 
internal structure reflecting charge rearrangement as we move in the 
moduli space, as in the previous subsection for the $SU(3)$ case.

Note that different regions in moduli space can also be described 
by field configurations that look quite different from the one above. 
For instance the dyon in the region in moduli space shown on the right 
side in Figure~\ref{websSU4} is best described by the 3-center field 
configuration essentially made of two $SU(3)$ webs 
$(2,1)-(-1,0)-(-1,-1)$ and $(0,1)-(-1,-1)-(1,0)$,
{\small
\bea
X_0 = {e\over |\vs-\vs_1|} + {e\over |\vs-\vs_2|} - X_0^0\ , \ \ 
Y_0 = {g\over |\vs-\vs_2|} - Y_0^0\ , &&
X_1 = -{e\over |\vs-\vs_1|} + X_1^0\ , \ \ Y_1 = Y_1^0\ ,  \nonumber\\
X_3 = {-e\over |\vs-\vs_2|} + {e\over |\vs-\vs_3|} + X_3^0\ , \ \ 
Y_3={-g\over |\vs-\vs_2|} + Y_3^0\ , && 
X_2 = -{e\over |\vs-\vs_3|} + X_2^0\ , \ \ Y_2=Y_2^0\ , \ \ 
\eea }
with the boundary conditions on the scalars near the cores being
\bea
\vs\ra\vs_1: && (X_0,Y_0)\ra ({X_1^0}',{Y_1^0}') \leftarrow (X_1,Y_1)\ ,
\nonumber\\
\vs\ra\vs_2: && (X_0,Y_0)\ra ({X_2^0}',{Y_2^0}') \leftarrow (X_3,Y_3)\ ,
\nonumber\\
\vs\ra\vs_3: && (X_3,Y_3)\ra ({X_3^0}',{Y_3^0}') \leftarrow (X_2,Y_2)\ .
\eea
It is straightforward to work out the core sizes and separations from 
these, and we find
\be
g\al_{12}=Y_0^0+Y_1^0\ , \qquad g\al_{23}=Y_3^0-Y_2^0\ ,
\ee
for the separations between the cores, whose charges now are
\be
\vs_1\equiv \{(1,0)_0 (-1,0)_1\}\ , \quad 
\vs_2\equiv \{(1,1)_0 (-1,-1)_3\}\ , \quad 
\vs_3\equiv \{(1,0)_3 (-1,0)_2\}\ ,
\ee
the asymptotic charges being the same of course. The constraint 
equation on the moduli values is the same as (\ref{modconst21SU4}).
This field configuration is reliable near different limits in moduli 
space, which can be analyzed as before.

Similar techniques can be used to study other dyons/webs in $SU(4)$.

\subsection{The \Nf\ SYM ``Halo''}

Now let us generalize the $(2,1)_0$ dyon in the previous section and 
consider the (maximally split)\ $(n,1)_0$ state, with one monopole, 
in an \Nf\ $SU(n+1)$ theory. The asymptotic charges of this state 
and those of the $n+1$ individual charge cores are
\bea
\{\ (n,1)_0\ (-1,0)_1\ \ldots\ (-1,0)_n\ (0,-1)_3\ \} \qquad\ 
{\rm at\ spatial\ infinity}\ ,\nonumber\\
\vs_{E_k}\equiv \{(1,0)_0 (-1,0)_k\}\ , \quad k=1,\ldots,n,\ \qquad\ 
\vs_M\equiv \{(0,1)_0 (0,-1)_1\}\ .
\eea
Then the constraint equations (\ref{nmConstraints})\ (specifically 
${Y_{E_k}^0}'$) give\ 
\be
g\al_{E_k,M}=Y_0^0+Y_k^0 \qquad \Rightarrow \qquad 
r_{E_k,M}={g\over Y_0^0+Y_k^0}\ , 
\ee
which represents the $n$ electric charges $E_k$ distributed on an 
approximate shell around the monopole $M$. For all $Y_k^0=Y_1^0$, this 
is an exact shell of radius ${g\over Y_0^0+Y_1^0}$.  This seems to be 
the \Nf\ field theory version of the ``halo'' in \Nt\ d=4 string 
theories of Denef \cite{denefHalo}. In the limit $Y_0^0,Y_k^0\ra 0$, 
all the lines of marginal stability coincide and all separations 
${1\over \al_{E_k,M}}$ grow, the ``halo'' size diverging.

Note that the separations between the electric charges is again not
fixed by the equations. Some intuition for this is gained by realizing
that if the monopole were not present, then the $n$ electric charges
are all ${1\over 2}$-BPS with unconstrained core separations since the
charge locations are all moduli. The presence of the monopole fixes
the electric-magnetic separations but does not affect the
electric-electric ones.

\subsection{More general $(n,m)_0$ dyons}

The structure of the $(n,m)_0$ state, \ie\ the core sizes and separations, 
with more than one monopole is harder to obtain in general since the 
equations are more complicated. We can see a pattern from the above 
general constraint equations (\ref{nmConstraints}), as well as the 
simpler examples of $SU(3), SU(4)$ webs described earlier. The $n+m$ 
equations given by ${X_{E_k}^0}', {Y_{M_k}^0}'$ are used to solve for 
the core sizes $\al_{E_k,E_k}, \al_{M_k,M_k}$, while we expect that 
the remaining $n+m$ equations (\ie\ ${X_{M_k}^0}', {Y_{E_k}^0}'$) which 
do not contain the core sizes can be used to solve for the core 
separations $\al_{E_i,M_j}, i\neq j$. However, clearly these $2(n+m)$ 
equations for the ${(n+m)(n+m+1)\over 2}$ unknown core sizes and 
separations\ $\{\al_{a_k,a_{k'}}\ , a_k=E_i,M_j\}$ are too few for an 
exact description of the spatial structure of the $(n,m)$ state on the
entire moduli space. In what follows, we will describe certain
limiting regions of the moduli space where we can in fact describe the
structure of the dyonic state in terms of approximate solutions to the
above equations.
This relies on the fact that a 2-center configuration near the LMS
acquires a loosely bound molecular structure where the microscopic
nonabelian physics of the charge cores can be ignored reliably. We can
thus look for regions in the moduli space where there is a $hierarchy$
of scales set up so that we first find a 2-center configuration near
one LMS, one of whose centers is itself made of two further centers near 
a LMS and so on. This sort of tree-like structure becomes increasingly 
more reliable in the vicinity of the simultaneous coincidence of all 
the lines of marginal stability, where all core separations are large.

\begin{figure}
\bc
\epsfig{file=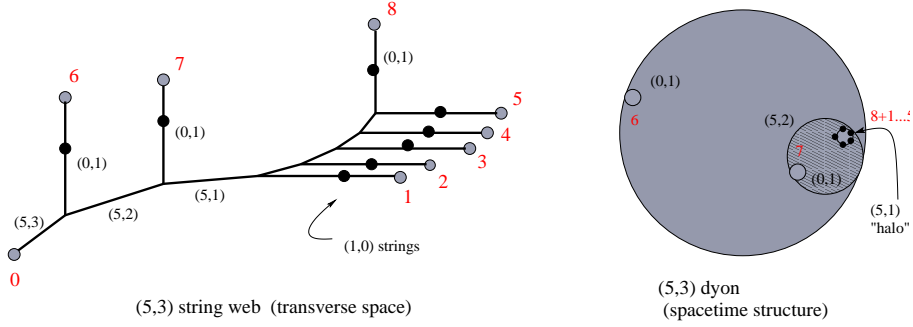, width=12cm}
\caption{{\small $(n,m)_0$ string webs illustrated using a $(5,3)_0$ 
example. The shape of the $(5,3)$ web in transverse space is shown, 
alongwith its approximate internal structure in spacetime (in this 
region of moduli space). In web: grey circles are D-branes, black 
circles are prong gluing locations.}}
\label{nmweb}
\ec
\end{figure}

We will illustrate this now with the example of the $(5,3)_0$ state. 
For convenience, let us label the $5$ electric charge centers
$E_k=1,\ldots,5$ and the $3$ magnetic charge centers $M_k=6,7,8$. The
constraint equations (\ref{nmConstraints}) then become:
\bea\label{53Constraints}
{X_1^0}'=e(\al_{11}+\al_{12}+\ldots +\al_{15})-X_0^0=-e\al_{11}+X_1^0\ , 
&& \quad \ldots , \quad {X_5^0}'=\ldots\ ,\quad  \nonumber\\
{Y_1^0}'=g(\al_{16}+\al_{17}+\al_{18})-Y_0^0=Y_1^0\ , 
&& \quad \ldots , \quad {Y_5^0}'= \ldots \ ,\quad \nonumber\\
{X_6^0}'=e(\al_{61}+\al_{62}+\ldots+\al_{65})-X_0^0=X_6^0=0\ , 
&& \quad \ldots \ , \quad {X_8^0}'= \ldots \ ,\quad \\
{Y_6^0}'=g(\al_{66}+\al_{67}+\al_{68})-Y_0^0=-g\al_{66}+Y_6^0\ ,
&& \quad \ldots , \quad {Y_8^0}'=\ldots \ . \nonumber
\eea
(For simplicity and ease of illustrating the physics, we have set $X_6^0=0$, 
without loss of generality.)
Solving these equations in general is hard. However, consider the region 
in moduli space where $X_0^0\ll X_7^0\ll X_8^0$. Then we can find the 
approximate solution of interest in an iterative way. First assume that 
there are only two centers of charge $M_6\equiv (0,1)$ and $D\equiv (5,2)$: 
\ie\ the centers $1\ldots 5, 7,8\equiv D$ constitute an effectively 
pointlike center of charge $(5,2)$. This gives the inverse core 
separation\ $e\al_{6D}={X_0^0\over 5}$, with appropriate inverse core 
sizes $\al_{66},\al_{DD}$. Now the dyon center $D\equiv (5,2)$ itself is 
not really pointlike of course but has structure, which can be obtained 
by treating $(5,2)$ as made of constituents $M_7\equiv (0,1)$ and 
$D'\equiv (5,1)$: this gives $e\al_{7D'}={X_7^0\over 5}$. The center 
$(5,1)$ of course we know to be a ``halo'' from above.\\
Since $X_0^0\ll X_7^0\ll X_8^0$, this zeroth order solution can thus 
be tweaked consistent with the full set of equations and we find the 
approximate solution (using the equations (\ref{53Constraints}) for 
${X_{6,7,8}^0}'$):
\bea\label{sep53}
&& e\al_{67},e\al_{68}\sim e\al_{61,}\ldots, e\al_{65} = {X_0^0\over 5}\ ,
\nonumber\\
&& e\al_{78}\sim e\al_{71},\ldots, e\al_{75} = {X_0^0+X_7^0\over 5}\ , 
\qquad\ \ e\al_{81},\ldots, e\al_{85} = {X_0^0+X_8^0\over 5}\ .
\eea
However, since these inverse core separations $\al_{E_i,M_j}$ between the 
electric and magnetic charges also appear in the equations for 
${Y_{1,\ldots,5}^0}'$, we must make sure that the above separations are 
consistent with the latter. Indeed substituting (\ref{sep53}) in 
the ${Y_{1,\ldots,5}^0}'$ equations (\ref{53Constraints}) gives:
\be
\al_{18} \sim {1\over g} (Y_0^0+Y_1^0) - {X_0^0\over 5e} 
- {X_0^0+X_7^0\over 5e}\ , 
\qquad \ldots
\ee
Adding these up and using the ${X_8^0}'$ equation, we get the nontrivial 
constraint for consistency of this set of inverse core separations:
\be\label{53constr}
{5e\over g}Y_0^0 - 3X_0^0 + {e\over g}\sum_{i=1}^5 Y_k^0 = X_7^0 + X_8^0\ .
\ee
To see that this is in fact sensible, note that this constraint equation 
reduces to\ ${5e\over g}Y_0^0-3X_0^0=0$, if $X_0^0, Y_0^0$ are large, 
which is the line in transverse space along which the single $(5,3)$ 
string, that the system reduces to, stretches (see Figure~\ref{nmweb}). 
Alternatively, for $X_0^0, Y_0^0$ small, the constraint reduces to\ 
${e\over g}\sum_{i=1}^5Y_k^0=X_7^0+X_8^0$. To understand the physics of 
this constraint, note that for all $Y_k^0=Y_1^0$ say, and $X_7^0\sim X_8^0$, 
this becomes\ $5eY_1^0=2gX_7^0$, representing the line in transverse space 
along which the $(5,2)$ string stretches\footnote{Note that the geometry 
of the web in transverse space forces $X_8^0\geq X_7^0$.}, while for 
$X_8^0\gg X_7^0$, we get\ $5eY_1^0=gX_8^0$, representing the profile of 
the $(5,1)$ string.

Finally, let us compare the core sizes with the core separations of 
relevance, to see the regimes of validity where the cores can be treated 
as pointlike and widely separated. From the ${X_{1,\ldots,5}^0}', 
{Y_{6,7,8}^0}'$ equations (\ref{53Constraints}), we get
\be\label{size53}
e\al_{E_i,E_i} = {1\over 2} (X_0^0+X_i^0 - \sum_{j\neq i} \al_{E_i,E_j})\ , 
\qquad 
g\al_{M_i,M_i} = {1\over 2} (Y_0^0+Y_i^0 - \sum_{j\neq i} \al_{M_i,M_j})\ .
\ee
For example, we have\ $\al_{ii}\sim {X_i^0\over 2e}\gg \al_{8i},\al_{7i},
\al_{6i}$ ,\ \ie\ center $E_i$ is pointlike (its inverse core size is much 
larger than its inverse separation from centers $6,7,8$) if\ $X_i^0\gg 
X_8^0,X_7^0,X_0^0$. Similarly, the magnetic charges can be regarded as 
pointlike compared with their relative separations if certain conditions 
hold on the vevs: \eg\ $\al_{66}\sim {Y_6^0\over 2g}\gg \al_{67},
\al_{68}$\ \ie\ charge $M_6$ is pointlike if\ $eY_6^0\gg gX_0^0$, while 
$\al_{77}\sim {1\over 2}({Y_7^0\over g}-{X_7^0\over 5e})\gg \al_{78}$\ if\ 
$Y_7^0\gg {3g\over 5e} X_7^0$, \ie\ the point $(X_7^0,Y_7^0)$ lies far 
above the line with slope ${3\over 5}$ (corresponding to the $(5,3)$ 
string). Likewise for charge $M_8$ to be pointlike, we have\ 
$\al_{88}\sim {1\over 2}({Y_8^0\over g}-{X_7^0\over 5e})\gg \al_{78}$\ 
if\ $Y_8^0\gg {3gX_7^0\over 5e}$ ,\ while\ $\al_{88}\gg\al_{8i}$\ if\ 
$eY_8^0\gg {g(X_7^0+2X_8^0)\over 5}$ (which implies the earlier 
condition). These conditions are compatible and give a region in moduli 
space where this molecular structure of the configuration is reliable. 

From Figure~\ref{nmweb}, we see that the limit $X_0^0,X_7^0,X_8^0 \ra 0$ 
(which through the constraint (\ref{53constr}) implies 
$Y_{1,\ldots,5}^0\ra 0$) corresponds to the various lines of marginal 
stability (\ie\ the various string junction locations) coinciding. In 
this limit, the inverse core separations (\ref{sep53}) all vanish, so 
that all charges are widely separated and the bound state unbinds. 
Keeping $X_{1,\ldots,5}^0,Y_{6,7,8}^0$  fixed in this limit ensures that 
the individual constituents with charges 
\be
\vs_{1,\ldots, 5}\equiv \{(1,0)_0 (-1,0)_{1,\ldots, 5}\}\ , \qquad 
\vs_{6,7,8}\equiv \{(0,1)_0 (0,-1)_{6,7,8}\}\ ,
\ee
are all pointlike to arbitrary accuracy, their core sizes 
$\al_{ij}={1\over r_{ij}}$ using (\ref{size53}) in this limit being
\bea
&& X_0^0\ll X_7^0\ll X_8^0 \ra 0\ , \qquad X_{1,\ldots,5}^0,Y_{6,7,8}^0\ 
{\rm fixed}: \nonumber\\
&& \al_{E_i,E_i} \sim {X_i^0\over 2e}\ ,\quad i=1,\ldots,5, \qquad\ \ 
\al_{M_j,M_j} \sim {Y_j^0\over 2e}\ ,\qquad j=6,7,8\ .
\eea

As we move away from this region, \eg\ as $X_0^0,Y_0^0$ grow relative
to the other vevs, the different centers begin to coalesce and their
spatial separations cannot be distinguished clearly enough from their
core sizes and the structure becomes fuzzy. Finally for large
$X_0^0,Y_0^0$, we see the single dyon center $(5,3)_0$. Clearly there 
exist different regions of moduli space where the $(5,3)_0$ can break 
up into different constituents. For instance, consider the region in 
moduli space 
\be
(X_5^0,Y_5^0)=(X_7^0,Y_7^0)\ , \qquad\ \ \ 
(X_8^0,Y_8^0)=(X_i^0,Y_i^0)=(X_1^0,Y_1^0)\ \quad i=1,2,3,4\ ,
\ee
where branes-$5,7$ and separately branes-$1,2,3,4,8$ coincide. Now 
the $(5,3)_0$ dyon can break up into four constituents resulting 
in an effectively $SU(4)$ string web \eg\
$(5,3)-(-4,1)-(-1,1)-(0,-1)$, where the $(1,1)$ and $(0,1)$ legs end
on branes-$7$ and -$6$ respectively, while the $(4,1)$ leg ends on
brane-$1$. The internal structure of the $(5,3)_0$ dyon in this region
can be analyzed by considering the appropriate limits in
(\ref{53Constraints}), resulting in a 3-center bound state
configuration somewhat similar to the $SU(4)$ dyons described earlier.

\section{More on internal structure transitions: string webs with 
internal faces}

\begin{figure}
\bc
\epsfig{file=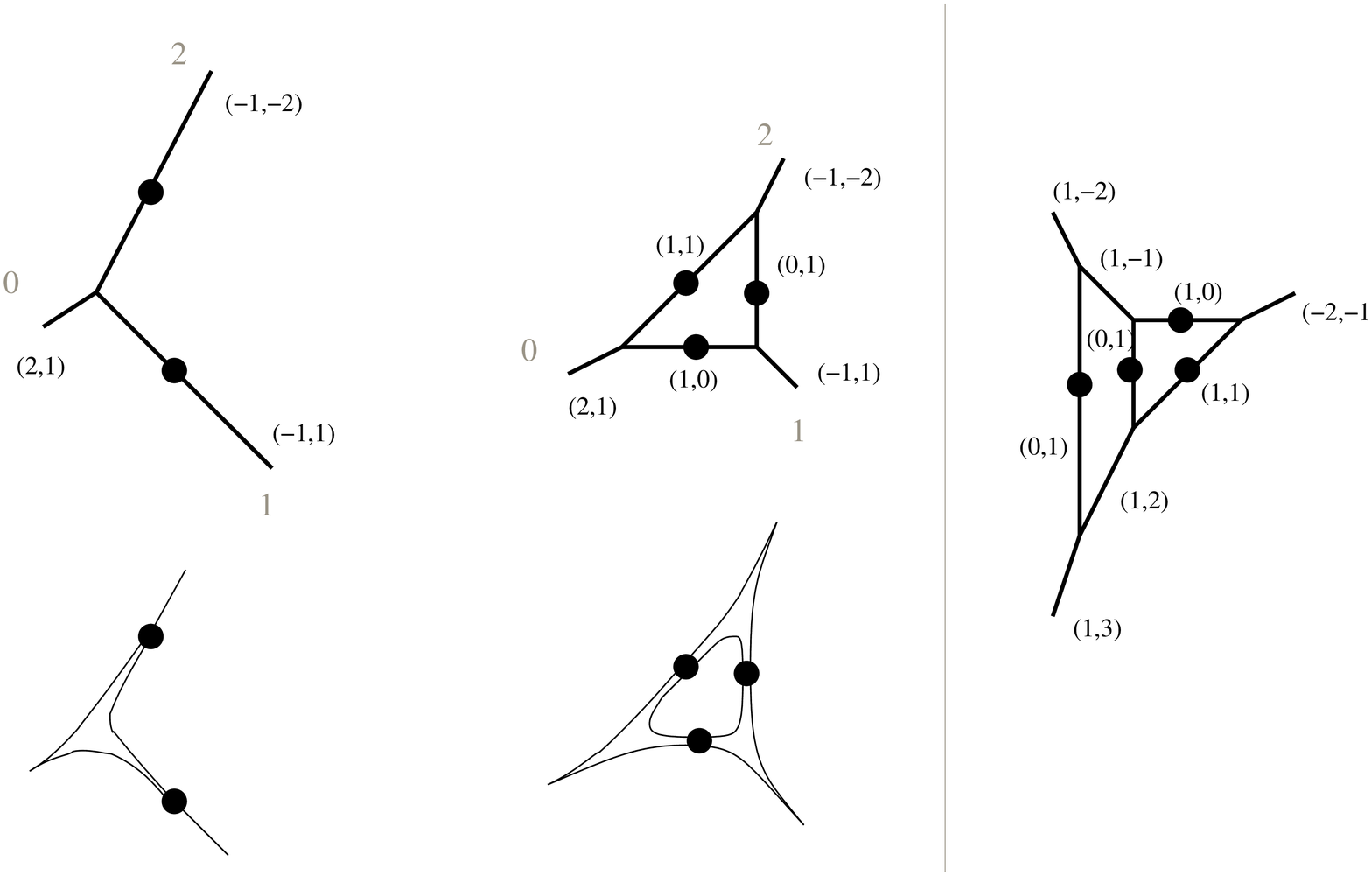, width=12cm}
\caption{{\small Webs with internal faces: on the left side is shown 
a web with a single internal face and its brane prong realization, while 
the right side shows a web with two internal faces (and the prong gluing 
locations). The charges of the various string legs are also shown.}}
\label{webfaces}
\ec
\end{figure}

It is known that there are moduli in string webs corresponding to
internal faces opening up (see \eg\ \cite{Bergman:1998gs, Kol:2000tw, 
Larus0101}). We will consider what is perhaps the simplest such string 
web\ $(2,1)-(-1,1)-(-1,-2)$\ in $SU(3)$ SYM theory (left side of
Figure~\ref{webfaces}).  As can be seen from the Figure, the internal
face opening up has internal string legs corresponding to $(1,1),
(1,0), (0,1)$ strings.

We would like to understand the spacetime structure of the dyon in
\Nf\ SYM theory that corresponds to such a web with an internal
face. Towards this end, we see that the field configuration
\bea\label{genus1}
X_0 = {e\over |\vs-\vs_1|} + {e\over |\vs-\vs_2|} - X_0^0\ , \quad\ 
Y_0 = {g\over |\vs-\vs_2|} - Y_0^0\ , && \quad
X_1 = -{e\over |\vs-\vs_1|} + X_1^0\ , \ \ \nonumber\\ 
X_2 = -{e\over |\vs-\vs_2|} + X_2^0\ , \quad\ 
Y_2=-{g\over |\vs-\vs_2|}-{g\over |\vs-\vs_3|}+Y_2^0\ , && \quad 
Y_1={g\over |\vs-\vs_3|}-Y_1^0\ , \ \
\eea
is a description of the web with internal faces\ (middle of 
Figure~\ref{webfaces}). This configuration consists of three $SU(3)$ 
3-pronged string web configurations, of charges\ $[(2,1)-(-1,0)-(-1,-1)],\ 
[(-1,1)-(1,0)-(0,-1)],\ [(-1,-2)-(1,1)-(0,1)]$,\ one pair 
$(X_i(\vs),Y_i(\vs))$ from each of branes-$0,1,2$,\ glued pairwise at 
appropriate charge cores. To elaborate, the spike from core $\vs_1$ in 
prong-$0$ glues onto that from $\vs_1$ in prong-$1$, while the 
$\vs_2$-spike in prong-$0$ glues onto that from $\vs_2$ in
prong-$2$. Finally the $\vs_3$-spikes from prongs-$1,2$ glue onto each
other. Alternatively, one could imagine branes-$0,1,2$ to have charge
cores $\{\vs_1,\vs_2\},\ \{\vs_3,\vs_4\},\ \{\vs_5,\vs_6\}$
respectively. Consistency of the prong-gluing then forces
$\vs_3\equiv\vs_1$,\ $\vs_2\equiv\vs_5$,\ and $\vs_4\equiv\vs_6$.

With the $(1,0)$ leg lying parallel to the $X$-axis, $Y={Y_1^0}'$, and 
scalar boundary conditions 
\bea
\vs\ra\vs_1: && (X_0,Y_0)\ra ({X_1^0}',{Y_1^0}') \leftarrow (X_1,Y_1)\ ,
\nonumber\\
\vs\ra\vs_2: && (X_0,Y_0)\ra ({X_2^0}',{Y_2^0}') \leftarrow (X_2,Y_2)\ ,
\nonumber\\
\vs\ra\vs_3: && (X_2,Y_2)\ra ({X_3^0}',{Y_3^0}') \leftarrow (X_3,Y_3)\ ,
\eea
it is straightforward to work out the core separations from the 
resulting constraint equations for the moduli as in our previous 
discussions, and we find
\be
g\al_{12}=Y_0^0+{Y_1^0}'\ , \qquad g\al_{13}=Y_1^0+{Y_1^0}'\ ,
\qquad e\al_{23}=X_2^0-X_1^0+{e\over g} (Y_1^0+{Y_1^0}')\ ,
\ee
with corresponding core sizes, ${Y_1^0}'$ being the modulus corresponding 
to changing the size of the internal face. We see that the core 
separations can be tuned independently showing the three lines of 
marginal stability. In the limit\ $Y_0^0+{Y_1^0}', Y_1^0+{Y_1^0}'\ra 0,\
X_2^0\ra X_1^0$, \ie\ short $(2,1), (-1,1), (-1,-2)$ web legs, all
core separations diverge and we obtain pointlike widely separated
cores, and a reliable field configuration. Unlike the $SU(4)$ 3-center 
configurations in Sec.~4.1, all core separations here are fixed and 
we have a rigid ``molecule'' of three constituent ``atoms'', shaped 
like an isosceles triangle, with sides\ 
$r_{13}=r_{23}={g\over Y_1^0+{Y_1^0}'}\ ,\ r_{12}={g\over Y_0^0+{Y_1^0}'}$ , 
for $X_2^0=X_1^0$.\ When the internal face is of maximal size, \ie\ 
zero size external web legs, we have\ ${Y_1^0}'=-Y_1^0$\ (and 
corresponding other edge locations, \eg\ ${X_3^0}'=X_1^0$):\ then we 
have\ $r_{12}={g\over Y_0^0-Y_1^0}$ and $r_{13}, r_{23}$ infinite, 
showing\ $Y_0^0\geq Y_1^0$ for a physical dyon configuration 
corresponding to a maximal size face. This then shows that the 
triangle inequality for the 3-center configuration is always satisfied 
and the molecule shape does not degenerate.

This configuration and the string web then show that the dyon has 
asymptotic charges and three individual constituent cores of charge
\bea
\{\ (2,1)_0\ (-1,1)_1\ (-1,-2)_2\ \} \qquad\ 
{\rm at\ spatial\ infinity}\ ,\nonumber\\
\vs_1\equiv \{(1,0)_0 (-1,0)_1\}\ , \quad 
\vs_2\equiv \{(1,1)_0 (-1,-1)_2\}\ , \quad 
\vs_3\equiv \{(0,1)_1 (0,-1)_2\}\ .
\eea
Thus each internal leg corresponds to a charge core.
By comparison, we see that the dyon for the same web but without the 
internal face corresponds to a 2-center configuration whose constituent 
cores have charges
\be
{\vs_1}'\equiv \{(1,-1)_0 (-1,1)_1\}\ , \qquad 
{\vs_2}'\equiv \{(1,2)_0 (-1,-2)_2\}\ ,
\ee
the corresponding scalar field configuration (with brane-$0$ near LMS) 
being
\bea\label{genus0}
X_0 = {e\over |\vs-{\vs_1}'|} + {e\over |\vs-{\vs_2}'|} - X_0^0\ , \quad &&
X_1 = -{e\over |\vs-{\vs_1}'|} + X_1^0\ , \quad 
Y_1={g\over |\vs-{\vs_1}'|}+Y_1^0\ , \nonumber\\ 
Y_0 = -{g\over |\vs-{\vs_1}'|}+{2g\over |\vs-{\vs_2}'|} - Y_0^0\ , \quad &&
X_2 = -{e\over |\vs-{\vs_2}'|} + X_2^0\ , \quad 
Y_2=-{2g\over |\vs-{\vs_2}'|}+Y_2^0\ , 
\eea
the prong-$0$ gluing onto spikes-$1,2$ at the black circles shown on the 
left side of Figure~\ref{webfaces}.

Thus we see that the spacetime description of the opening up of the
internal face is essentially similar to the kinds of transitions in
internal structure that we saw earlier in Sec.~3. However, in this
case, we are not moving around in the moduli space of the D-branes
defining the vacuum: the D-brane locations are fixed. Changing moduli
of the internal structure gives rise to charge rearrangement of the
cores. Specifically we go to the symmetric point in moduli space of
the 2-core configuration (mentioned at the end of Sec.~2) which 
effectively has a single coalesced core: this then breaks up into 
the 3-core configuration with rearranged charges.

The way the brane prong configurations glue onto each other is shown
in Figure~\ref{webfaces}\ (the thickened webs on the bottom), the
black circles being the gluing locations (the grey circles
corresponding to D-brane locations are not shown explicitly). A
noteworthy point is that this system exhibits change in topology of
the branes' collective worldvolume 3-surface. With no internal face,
we have the 3-dimensional analog of a genus-0 surface (\ref{genus0}),
while the opening up of the internal face creates a new ``handle'' in
the interior of the brane 3-surface (\ref{genus1}). This ``tearing'' of
the 3-surface as we transit from no handle to one handle would seem
like a singular process from the point of view of classical geometry,
but we have to remember that our abelian approximations are not good
in this intermediate region. The process of separation of the new
cores from the single ``effective'' core involves the light nonabelian
fields that we have neglected. We expect that the full nonabelian
theory gives a smooth description of this brane prong topology change
as the internal face opens up.

It is possible to describe webs with multiple internal faces along 
these lines. The right side of Figure~\ref{webfaces} shows the web\ 
$(1,3)-(-2,-1)-(1,-2)$, with two internal faces opened up. We can 
construct a field configuration for this system by considering the 
two branes with legs $(1,3), (1,-2)$ to have a 4-pronged web scalar 
configuration each (of the sort described earlier in Sec.~4), and 
the brane with leg $(-2,-1)$ to be a 3-pronged web. These then glue 
onto each other (at the black circles) as shown in the Figure.

\section{Discussion}

We have described field configurations in an abelian $U(1)^r$ 
approximation that describe dyons and their approximate internal 
structure in \Nf\ $SU(N)$ super Yang-Mills theory. These dovetail 
closely with string web configurations corresponding to the dyon 
states in D-brane constructions of these gauge theories. The internal
structure of the dyon for fixed asymptotic charges is a complicated 
function of the moduli space in general, as we have seen, closely 
intertwining with the geometry of the corresponding string web in 
transverse space.

One could now ask how reliable these constructions are, in terms of
giving insight into the dyon internal structure. Note that while the
$SU(3)$ dyon/web is truly ${1\over 4}$-BPS, the generic such state in
a higher rank $SU(N)$ theory is not, for the following
reason\footnote{This arose in discussions with Ashoke Sen.}.  Recall
that the BPS bound equations follow from restricting to a
2-dimensional subspace of the 6N-dimensional Coulomb branch: this
corresponds to the fact that the generic string web is a planar
configuration stretched between $N$ D-branes. This is crucial to
having ${1\over 4}$-th supersymmetry preserved. However a small
perturbation to the locations of one or more of the D-branes will tilt
the string web which will now no longer be planar: from the field
theory point of view per se, this means one of the other $4N$
transverse scalars has been turned on, ruining the Bogomolny bound in
the energy functional. Thus the dyon/web is no longer precisely
${1\over 4}$-BPS, with the exception being $SU(3)$\ (since 3 points
always lie on some plane)\footnote{See \eg\ \cite{Dyons07, atish0702150, 
sen2ctr0705} for some related discussions on lines of marginal 
stability in \Nf\ string theories.}.

From the field theory point of view, we have $k$ charge centers, held
together in equilibrium by force balance\ (approximately, since some
of the core separations are not determined by the methods here).  In
the regions of moduli space we have focussed on, \ie\ near one or more
lines of marginal stability, the charge cores are all effectively
pointlike and widely separated. Thus it seems reasonable to suppose
that the internal structure in spacetime that we have obtained here is
in fact not drastically altered even though the dyon/web is not truly
${1\over 4}$-BPS, at least at weak coupling.  Essentially any such
dyon is approximately a classical object in this regime, being a
molecule-like configuration of widely separated pointlike charge
cores, so that quantum corrections would seem to make negligible
contributions to the force balance conditions.  This corroborates with
the fact that at weak string coupling, the string web is a relatively
light object stretched between heavy D3-branes.  Perhaps this makes
the picture here more interesting since we seem to be describing
approximate configurations, robust under small perturbations, for
highly nontrivial non-supersymmetric bound states in \Nf\ super
Yang-Mills theories of high rank. It would be very interesting to go
beyond the abelian approximation here, perhaps by adding higher
derivative terms to the effective action, or by considering the
corresponding configurations in the full nonabelian theory.

As we have seen, the boundary conditions on the scalar moduli 
generically fix the electric-magnetic core separations, or more 
generally, those of constituents with mutually nonlocal charges. 
However the separations of mutually local charges are not fixed, so 
these must be regarded as moduli of the configurations. Thus the low 
energy dynamics of these dyons would be captured by an appropriate 
approximately supersymmetric quantum mechanics on the moduli space 
of these solutions. There presumably are similar moduli for dyon 
states that do not satisfy $n,m$ prime: in this case, some 
constituent cores would perhaps fragment. A systematic study, 
possibly equipped with a better understanding of the coupling 
dependence of these configurations, might draw connections 
to the quiver quantum mechanics of \cite{denefHalo}.

Finally, it would be interesting to understand black hole bound state
configurations, in part along the lines described in \cite{denef,
  denefHalo}, and more recently \cite{denefmoore, atish0702150,
  sen2ctr0705} (see also \eg\ the review \cite{senRevAttr07}), and
contrast them with the description of SYM dyons here. For instance,
away from a line of marginal stability, we have seen that the SYM dyon
centers here begin to coalesce, and nonabelian modes (\ie\ the
ultraviolet completion of the abelian approximation) become
important. It would be interesting to understand if similar
ultraviolet completions, \ie\ stringy corrections to the low energy
gravity description, are required for a careful understanding of the
internal structure of black hole bound states away from lines of
marginal stability (which would naively seem to involve merging or
bifurcation of horizons).  On a more technical note, perhaps some of
these spacetime configurations for dyons corresponding to webs with
internal faces might be of relevance for understanding black holes in
Type IIB compactifications on $K3\times T^2$, via effective string
webs wrapped on $T^2$ corresponding to higher genus surfaces
\cite{gaiotto, atish0702150}.

More generally, this approach would perhaps closely tie into the broad
ideas and attempts to understand black holes by decomposing it into
constituent bits \cite{BHbits}.

\vspace{10mm}
{\bf Acknowledgements}: It is a pleasure to thank Frederik Denef and
Ashoke Sen for discussions at the early stages and towards the
completion of this work respectively. I have also benefitted from
discussions with Atish Dabholkar and Suresh Nampuri on analyzing the
degeneracies of Stern-Yi string web states from a string theory point
of view. I'd like to thank the Organizers of the ``From Strings to the
LHC: II'' school+conference, Bangalore, India, for hospitality during
the final stages of this work.

%\vspace{8mm}
\appendix
\section{BPS bounds, brane prongs and string webs}

Here we review the BPS state construction \cite{Argyres:2001pv,
  Argyres:2000xs} of string webs from the low energy effective action
on the Coulomb branch, making perhaps slightly more manifest 
the brane prong description, paving the way for the configurations
described in this paper. First, we consider an \Nf\ $U(1)$ theory
($\vE,\vB$ being electric, magnetic fields) with two scalars $X,Y$
representing a 2-dim. subspace of the 6-dim. Coulomb branch.
With a view to finding BPS solutions, we complete squares in the 
energy functional to obtain Bogomolny bound equations: this gives 
{\small \bea
&& M = {1\over g_{YM}^2} \int d^3\vx\, {1\over2}\left[(\vE-\cos\alpha\vn X 
+ \sin\alpha\vn Y)^2 + (\vB-\sin\alpha\vn X - \cos\alpha\vn Y)^2\right]
\nonumber \\
&& {} \qquad\ \ + {1\over g_{YM}^2} \sum_{I=0}^n \left\{
\cos\alpha \oint_{S^2_I} (X\vE+Y\vB)\cdot d{\vec a} + 
\sin\alpha\oint_{S^2_I} (X\vB-Y\vE)\cdot d{\vec a}\right\}
\eea }
where the boundaries $S^2_I$ are spheres around each of several charge 
cores in the system and one sphere at infinity, and we have used the 
divergence-free equations for the electric and magnetic fields 
$\vn\cdot\vE=\vn\cdot\vB=0$ away from the cores\footnote{A similar 
Bogomolny bound in a nonabelian gauge theory would contain \eg\ a BPS 
't Hooft-Polyakov monopole solution, with the nonabelian modes dying 
out exponentially outside the monopole core whose size is set by the 
Higgs vev. The abelian description here can be regarded as the 
approximate Dirac-monopole-like (semiclassical) description outside 
the charge core.}. Labelling the 
boundaries so that the $I=0$ boundary is the one at infinity and the
$I=i\neq0$ are the ones around the charge cores, within the abelian
approximation, the scalars $X,Y$ can be regarded as taking constant
values $X^i,Y^i$ at the $i$-th boundary while at infinity they take
their asymptotic vacuum values, say $(-X^0,-Y^0)$. This reduces the 
boundary terms to expressions involving the electric/magnetic charges\
$\oint_{S^2_i}\vE\cdot d{\vec a}=Q_E^i\ ,
\ \oint_{S^2_i}\vB\cdot d{\vec a}=Q_B^i$, 
of the various cores. By charge conservation, the charges at infinity 
are $(Q_E^0,Q_B^0) = -\sum_{i=1}^n (Q_E^i,Q_B^i)$. Then the BPS 
saturated mass (using\ $g_s=g_{YM}^2$ in 4-dim)
\be
M_{BPS} = {1\over g_s}\ \sqrt{[(X^i+X^0)Q_E^i + (Y^i+Y^0)Q_B^i]^2 
+ [(X^i+X^0) Q_B^i - (Y^i+Y^0) Q_E^i]^2}\ ,
\ee
(implied sum over the $i$ charge cores) arising from the boundary 
terms, is obtained when the BPS bound equations
\be
\vE = \cos \alpha \vn X - \sin \alpha \vn Y\ , \qquad
\vB = \sin \alpha \vn X + \cos \alpha \vn Y\ ,
\ee
hold, where we maximize w.r.t. $\al$, \ie\ when
\be\label{tanal}
\tan\al = {(X^i+X^0) Q_B^i - (Y^i+Y^0) Q_E^i\over 
(X^i+X^0)Q_E^i + (Y^i+Y^0)Q_B^i}\ .
\ee
The scalars are harmonic, \ie\ $\nabla^2 X=\nabla^2 Y=0$. 
Let us illustrate this with a few examples. A single 
$(Q_E^1,Q_B^1)=(ne,mg)$ charge core has\ $\tan\al={Q_B^1\over Q_E^1}$, 
where $e,g$ are the units of electric and magnetic charge. A unit 
electric charge $Q_E^1=e , Q_B^1=0$, is then given by
\be
\vE = e {\vs-\vs_0\over 4\pi |\vs-\vs_0|^3}\ , \qquad
X = {ne\over 4\pi |\vs-\vs_0|} - X^0\ ,
\ee
where we have chosen $\al+\pi$ (which also satisfies (\ref{tanal})) 
so that the spike stretches along increasing $X$. This is a 
${1\over 2}$-BPS state.
Here we have set $Y^i=-Y^0$, \ie\ we have turned off the $Y$ scalar. 
Alternatively we have performed a rotation in the $(X,Y)$-plane so 
that the $(n,m)$-string emanating from a D-brane stretches along 
the $X$-axis. The $4\pi$ factors are not important for our analysis, so 
we will drop this for convenience.

Now consider a 2-center configuration with\ $Q_E^1=ne ,\ Q_B^2=mg ,\ 
Q_E^2=Q_B^1=0$, \ie\ one charge core is purely electric while the other 
is purely magnetic. This is a ${1\over 4}$-BPS state. This gives
\be
\vE=ne {\vs-\vs_1\over|\vs-\vs_1|^3}\ , \quad 
\vB=mg {\vs-\vs_2\over|\vs-\vs_2|^3}\ , \qquad\ \
\tan\al={mg X^0-ne Y^0\over (X^1+X^0)n+(Y^2+Y^0)m}\ ,
\ee
where we have imposed boundary conditions so that the moduli values 
at the charge cores are $(X,Y)_{\vs_1}\equiv (X^1,0),\ 
(X,Y)_{\vs_2}\equiv (0,Y^2)$, for simplicity. Now we see that\ 
$\al=0,\pi$\ if \ $mg X^0=ne Y^0$. Choosing $\al=\pi$, we have
$\vE=\vn X , \ \vB=\vn Y$, giving the scalar field configurations 
\be\label{prongU1}
X = {ne\over |\vs-\vs_1|} - X^0\ , \qquad Y = {mg\over |\vs-\vs_2|} - Y^0\ ,
\ee
the constants of integration being fixed to be the vacuum moduli values.
Effectively, our choice of $\al$ has fixed our freedom to perform a
rotation in the $(X,Y)$-plane, and ensured that \eg\ electric charges
correspond to F-strings stretched along the $X$-axis, and more
generally an $(n,m)$ charge corresponds to an $(n,m)$-string stretched
along a line of slope ${Y^0\over X^0}={m\over n} {g\over e}$ in the
$(X,Y)$-plane.

The scalar configurations can now be interpreted as brane prongs 
approximating a string web with increasing accuracy as we approach the 
line of marginal stability, which lies at $X^0,Y^0=0$ with our choices 
of moduli values.

Now let us go to higher rank theories. We expect on physical grounds 
that the $U(1)$ theory above is UV incomplete, and must be regarded as 
a piece of a $U(1)^n$ theory, arising as the Higgsed approximation 
to a $U(n)$ theory. The energy functional for the $U(1)^n$ theory is
\be
M = \int {\cal E} = {1\over2} \int \sum_i \left[ \vE_i^2 + \vB_i^2 + 
(\nabla X_i)^2 + (\nabla Y_i)^2 \right]
\ee
Extremizing this and finding BPS states is straightforward but not 
simple. For the $SU(3)$ theory, details appear in \cite{Argyres:2001pv,
 Argyres:2000xs}. We will simply describe some intuitive aspects of 
the embedding of the $U(1)$ brane-prong configurations into $U(1)^2$ 
and $U(1)^3$ theories: this should pave the way for the description 
of dyon-web configurations in the text.

We expect that a $U(1)$ brane-spike representing a single charge 
must be patched up with a corresponding spike from another brane, the 
two spikes together being thought of as a low energy $U(1)^2$ 
approximation to the full nonabelian $U(2)$ theory. Thus for, say,
a unit electric charge represented by a $(1,0)$-string stretched 
between two D-branes at say $(-X_0,0)$ and $(L,0)$ on the $(X,Y)$-plane, 
we expect the BPS bound equations 
\be
\vE_1 = \nabla X_1\ , \qquad \vE_2 = \nabla X_2\ ,
\ee
from the two D-branes separately, giving the scalar field configurations 
(\ref{spikesU12}).
For the two spikes to be glued together, we require that the two cores 
be located at the same position $\vs_0$ along the D-brane worldvolume 
directions and the core sizes be the same (within this approximation). 
This implies that the gluing happens at the midpoint of the line joining 
the two branes, \ie\ at the singularity corresponding to enhanced $U(2)$ 
gauge symmetry on the moduli space. Note that the gluing and charge 
boundary conditions are consistent with the fact that the $X_1$ brane 
appears to have a positive charge flux emanating from the gluing core,
while the $X_2$ brane appears to have a negative charge flux emanating
from the gluing location.

Similarly, the brane-prong configuration (\ref{prongU1}) representing 
say the $(1,1)-(-1,0)-(0,-1)$ string web in the $U(1)$ theory must be 
patched up with corresponding spikes/prongs from two other branes, the 
three spike/prong pieces together being thought of as a low energy 
$U(1)^3$ approximation to the full nonabelian $U(3)$ theory. Thus for, 
say, the $(1,1)-(-1,0)-(0,-1)$ string web stretched between three 
D-branes at say $0\equiv (-X_0,-Y_0), 1\equiv (X_1^0,0), 2\equiv 
(0,Y_2^0)$ on the $(X,Y)$-plane\ (Figure~\ref{webprongs}), we expect 
the $U(1)^3$ BPS bound equations 
\be
\vE_0 = \nabla X_0\ , \qquad \vB_0 = \nabla Y_0\ , \qquad\ \ 
\vE_1 = \nabla X_1\ , \qquad \vB_2 = \nabla Y_2\ .
\ee
The expectation that the electric part of prong-0 glues onto that of 
brane-1 while the magnetic part of prong-0 glues onto that of brane-2 
implies 
\be
\vE_0=e {\vs-\vs_1\over|\vs-\vs_1|^3}\ , \quad 
\vB_0=g {\vs-\vs_2\over|\vs-\vs_2|^3}\ , \qquad\ \
\vE_1=-e {\vs-\vs_1\over|\vs-\vs_1|^3}\ , \quad 
\vB_2=-g {\vs-\vs_2\over|\vs-\vs_2|^3}\ .
\ee
These field strengths alongwith the BPS bound equations imply the 
scalar field configurations in (\ref{111001web}).

It is similarly possible to write out educated guesses for the BPS 
bound equations for the higher rank/charge cases and thence the 
corresponding field configurations: \eg\ the BPS bound equations for 
the $(5,3)_0$ state described in the text are
\be
\vE_0 = \nabla X_0\ , \qquad \vB_0 = \nabla Y_0\ , \qquad\ \ 
\vE_{1,\ldots,5} = \nabla X_{1,\ldots,5}\ , \qquad 
\vB_{6,7,8} = \nabla Y_{6,7,8}\ ,
\ee
from which we can write the scalar field configurations (\ref{nmEqns}) 
specializing to the $(5,3)_0$ state.

The field configurations in the text are written with a rotation in
the transverse $(X,Y)$-plane, as for the single $U(1)$ theory
described earlier, so that the $(n,m)_0$ dyon spike from brane-$0$
(away from any line of marginal stability) points in the
$(X^0,Y^0)\sim (n,m)$ direction.

%\newpage
%\vspace{10mm}

\end{document}